\documentclass[12pt]{article}

\usepackage{latexsym}

\usepackage[dvipdfm]{graphicx}

\def\CI{Copenhagen interpretation }
\def\ci{Copenhagen interpretation}
\def\tr{{\rm tr}}
\def\ket#1{\mid~\!\!\!{#1}~\!\!\rangle}
\def\bra#1{\langle~\!\!{#1}~\!\!\!\mid}

\def\cN{{\cal N}}
\def\cD{{\cal D}}

\def\Q{quantum }
\def\QM{quantum mechanics }
\def\qm{quantum mechanics}
\def\QMl{quantum mechanical }
\def\qml{quantum mechanical}
\def\M{measurement }
\def\m{measurement}
\def\${\enskip$}

\def\I{interpretation }

\begin{document}

{\bf \large \noindent How Can the No-Collapse and\\ the Collapse
Interpretations of Quantum-Mechanics\\ Give the Same Description?}\\

\vspace{0.5cm}

{\bf \noindent F. Herbut}\\

\vspace{0.5cm}

\rm

\noindent {\bf Abstract} It is shown that no-collapse and collapse interpretations of \QM give equal object states (which predict everything that is observable) if one bases the relevant relations on the Von Neumann-L\"{u}ders 'projection'. This connection is elaborated in detail from simple to most general cases. Distinguishability of the two approaches, which exists in principle, is also discussed. In the very simple illustration of passing one slit physical-insight difficulties in the collapse approach are indicated. For the purpose of interference between the two wavefunction components also the Maxh-Zehnder interferometer is discussed.\\

\noindent {\bf Keywords} Interpretation; collapse; no-collapse; Everett theory; relative-state; many worlds; general relations\\

\vspace{1.5cm}

{\footnotesize \rm \noindent
\rule[0mm]{4.62cm}{0.5mm}

\noindent F. Herbut\\
Serbian Academy of Sciences and
Arts, Knez Mihajlova 35, 11000
Belgrade, Serbia\\
e-mail: fedorh@sanu.ac.rs}\\

\pagebreak

{\bf \noindent 1 Introduction}\\
There is a unique (first-quantization non-relativistic) \QMl formalism (the textbook material), but a diversity of interpretations of how it should become the branch of physics called \qm . One of the baffling features in this variety is the fact that there are collapse (reduction of the wave function, objectification) and no-collapse interpretations. We explore the question from the title in this investigation.

To begin with, it must be clarified {\bf how far} we {\bf assume} that \QM is {\bf valid}. According to Bohr's views and the Copenhagen \I \cite{Copenh}, \QM is mostly assumed to be applicable only to microscopic physical systems, and macroscopic systems are thought to be governed by classical physics. In contrast, in this investigation the validity of \QM is extended also to classical systems including human observers following Everett \cite{Everett1}, \cite{Everett2} and Cooper et al. \cite{Cooper}. If this so-called \textbf{extended \QM} comprises the whole world, then one speaks of \QM of universal validity. But in this study we confine ourselves to micro- and macroscopic systems including detectors.

Stipulation of extended \QM is a serious step away from the \ci , which was very successful for a long time. The \CI seemed so natural that it appeared to be the right \textbf{phenomenological physical use} of the \QMl formalism. One must have good experimental reason for considering any other \I than the Copenhagen one. (My late teacher Rudolf E. Peierls used to say: "I do not like the expression "\ci " because it suggests that there exists some other interpretation.")

The present author found experimental reason for replacing the \CI by extended \QM when he studied two extremely thought-provoking experiments of Scully et al. \cite{Scully1}, \cite{Scully2}. The authors of these remarkable articles
utilized the \QMl formalism only as a "book-keeping device" to calculate probabilities (which they experimentally confirmed). Thus, in the hands of these authors the \CI did not only restrict the validity of \QM to microscopic objects (photons in this case); it degraded the formalism to mere 'cooking recipe' for probabilities.

The most challenging idea in these experiments was Scully's  delayed-choice: one has a pair of photons propagating in opposite directions. One of them is absorbed by a detector, and only after this absorbtion has taken place it is decided on its 'partner' (automatically and in a random way) if the former photon was a which-way one (in the essentially double-slit experiment performed) or an interference one. Attempts to think about the fate of the photons in a collapse way seemed to suggest an effect going backwards in time. Hence, the \QMl formalism seemed doomed to the mentioned degradation.

The present author accepted the above-mentioned extended \QM of Everett and Cooper et al. in which the detectors were treated on the same footing as the photons. The result was a satisfactory \textbf{physical insight} in these, at first, perplexing experiments in terms of the first-quantization \QMl formalism \cite{FHScully1}, \cite{FHScully2}.

Next we turn to Niels Bohr's attitude towards extended \qm .\\

\pagebreak

{\bf \noindent 2 Bohr's macroscopic complementarity principle}\\
Shimony \cite{Shimony} (pp.769 and 770) wrote:

"I suspect that Bohr was aware of the difficulties inherent in a macroscopical ontology, and in his most careful writing he states subtle qualifications concerning states of macroscopic objects. For example
\cite{Bohr1},

\begin{quote}
"The main point here is the distinction between the {\it objects} under investigation and the {\it measuring instruments} which serve to define, in classical terms, the conditions under which the phenomena appear. Incidentally, we remark that, for the illustration of the preceding considerations, it is not relevant that experiments involving an accurate control of the momentum or energy  transfer from atomic particles to heavy bodies like diaphragms and shutters would be very difficult to perform, if practicable at all. It is only decisive that, in contrast to the proper measuring instruments, these bodies together with the particles would constitute the system to which the \QMl formalism has to be applied."
\end{quote}

Shimony further comments: "Bohr is saying that from one point of view the apparatus is described classically and from another, mutually exclusive point of view, it is described \qml ly. In other words, he is applying the principle of complementarity, which was originally formulated for microscopical phenomena, to a macroscopic piece of apparatus. ..."\\

It seems to me that in Bohr's macroscopic complementarity principle (as we may call it leaning on Shimony's discussion) there is concealed an important implication. If the \QMl formalism can be applied to
"bodies" like diaphragms and shutters, then it can, most likely, be applied to any classical systems. But then, clearly, Bohr admits that macroscopic systems, usually described classically, can be described also \qml ly. This actually opens the door for \QMl description of classical objects as it is often done nowadays \cite{ZehBOOK}, \cite{ZurekRMP}. Thus, seeds of modern results are present already in the "most careful writing" (as Shimony puts it) of the father of the \ci . At the very least, Bohr's quote seems to be in agreement with extended \QM as utilized in the present article.\\

Let us make straightforward use of the \QMl formalism to enable us to take up discussion of experiments.\\

{\bf \noindent 3 Same Description in CQM and in RSQM}\\
Returning to our aim, in any (Copenhagen-inspired) \textbf{collapse \I} of \QM (CQM) one may assume collapse in the form of occurrence of an event (projector) \$Q_2\$ of the second subsystem in a bipartite system in a pure state \$\ket{\Psi}_{12}\$.

To start with the simplest, we assume that the event \textbf{occurs in an ideal way, }i.e., as if it were ideally measured. (We generalize this assumption in subsection 6.1 .) Then the well-known L\"{u}ders selective-\M change-of-state relation \cite{Lud}, \cite{Messiah},  \cite{Laloe}, which has come to be known as the (basic form of the) Von Neumann-L\"{u}ders 'projection', reads $$ \ket{\Psi}_{12}\quad\rightarrow\quad\ket{\Psi}_{12}^c
 \equiv  Q_2\ket{\Psi}_{12}\Big/ \Big(\bra{\Psi}_{12}Q_2\ket{\Psi}_{12}\Big)^{1/2}.
 \eqno{(1)}$$  Here the arrow denotes collapse. It is easily seen that relation (1) is a generalized form of the famous projection postulate of von Neumann encompassing events expressed by projectors more general than ray projectors (elementary events).

  Since \$Q_2\$ is a subsystem event, one might intuitively expect that its occurrence has only local effects. Actually relation (1) shows that the change-of-state is global. In particular, it causes, in general, a change in the state of the other, distant (or remote)  subsystem, in our case subsystem \$1\$. (The term "distant" only expresses lack of dynamical influence, i. e., the fact that the influence is due only to the existing quantum correlations. There need not be spatial distance involved.) The distant-subsystem change-of-state (which is occasionally drastic) reads: $$\rho_1\equiv\tr_2\Big(\ket{\Psi}_{12}
 \bra{\Psi}_{12}
 \Big)\quad\rightarrow\quad \rho_1^c\equiv \tr_2\Big(\ket{\Psi}_{12}^c\bra{\Psi}_{12}^c
 \Big),\eqno{(2)}$$ where \$\tr_2\$ is the partial trace over subsystem \$2\$.

No-collapse \I of \QM was introduced by Everett \cite{Everett1}, \cite{Everett2}, and it is called the relative-state \I of \qm , or by acronym \textbf{RSQM}. In this approach to \QM one envisages a cut between subsystems \$1\$, which is to be the object of description, and subsystem \$2\$, which is left out from the object of description. Further, one tries to find a general relation giving the relative-state density operator \$\rho_1^{rs}\$, the counterpart of  (2), with a suitable so-called subject entity.

As one can see from (2) and (1), one can write $$\rho_1^c\equiv\tr_2\Big(\ket{\Psi}_{12}^c
\bra{\Psi}_{12}^c\Big)\equiv
 \tr_2\Big(Q_2(\ket{\Psi}_{12}\bra{\Psi}_{12})
 Q_2\Big)
 \Big/\Big(\bra{\Psi}_{12}Q_2\ket{\Psi}_{12}\Big).
 \eqno{(3a)}$$ Commutation under  a partial trace is valid if one of the two operators, like \$Q_2\$ in our case, acts in the factor space over which the partial trace is taken (as easily checked). One can then use this fact and idempotency of the projector to obtain the equivalent relation $$\rho_1^c=\tr_2\Big[\Big (\ket{\Psi}_{12}\bra{\Psi}_{12}\Big)Q_2\Big]\Big/ \Big\{ \tr\Big[\Big (\ket{\Psi}_{12}\bra{\Psi}_{12}\Big)Q_2\Big]\Big\} .\eqno{(3b)}$$

 Thus, if one takes {\bf by definition} (3b) for \textbf{the general relation} in RSQM, then $$\rho_1^{rs}=\rho_1^c,
 \eqno{(4)}$$ i. e., \textbf{the same description} by \$\rho_1^{rs}\$ is achieved in RSQM as by \$\rho_1^c\$ in \textbf{CQM}
 (collapse \qm ). By this the \textbf{subject entity} is the \textbf{subject event} \$Q_2\$ for subsystem \$2\$. (Note that the the subject entity consists of the choice of a subsystem and of an event on it. The former is displayed by the index on the projector.)\\

To justify taking (3b) for the general relation in RSQM we are obliged to show that what is known as Everett's original relation is a special case of (3b). This is accomplished in Appendices A, B, and C.\\

We are now prepared to investigate how the derived 'same description' in  RSQM and in CQM works in concrete experiments.\\

{\bf \noindent 4. One-Slit Preparation}

We begin illustration with \textbf{one-slit preparation} of some experiment. Experiments consist of preparation, evolution and \m . We investigate if the descriptions in CQM and in RSQM coincide. The discussion will be restricted to the preparation because it will prove unnecessary to extend it to evolution and \m .

The first subsystem is a particle or a photon, we'll say a quanton, the second is the screen. We think of
the screen as of an infinite
surface perpendicular to the
motion of the incoming quanton.
\pagebreak

{\bf \noindent  4.1 Version with a
Detected Triggering Event}

The screen is thought of as broken
up into two non-overlapping
segments: the slit is one of them
(segment \$s\$) and the rest of the
screen is the other (segment
\$\bar s\$). Being hit by the quanton in the latter, i.e., transfer of linear momentum at
this segment, corresponds to the
occurrence of, say, the event (projector)
\$Q^{\bar s}_2\$ on the screen.

Let us think of a so-called {\it negative-result \M } \cite{Renninger} consisting in the arrival of the quanton at the screen and the non-occurrence of \$Q^{\bar s}_2\$. This amounts to passing the slit. More precisely, on the screen one has {\it ideal occurrence} of the opposite event
$$Q^s_2=I_2-Q^{\bar s}_2,\eqno{(5)}$$ where \$I_2\$ is the identity operator (certain event) on the second subsystem.

For simplicity, we assume that the
composite-system \$(1+2)\$ is in a pure state \$\ket{\Phi ,t_i}_{12}\$ at the end of the interaction. We further assume that both \$Q^{\bar s}_2\$ and \$Q^s_2\$ have positive probabilities in
\$\ket{\Phi ,t_i}_{12}\$. By \$t_i\$ is denoted the initial moment of the experiment, i. e., the final moment of the preparation. One may refer to
the event \$Q_2^s\$ as to the {\bf
triggering event} of the preparation \cite{FHpreparARX}.

Applying \$Q_2^s+Q_2^{\bar s}\enskip \Big(=I_2\Big)\$
to the composite state, one obtains its relevant decomposition $$\ket{\Phi ,t_i}_{12}=\Big(\bra{\Phi ,t_i}_{12}Q_2^s
\ket{\Phi ,t_i}_{12}\Big)^{1/2}\Big[
Q_2^s\ket{\Phi ,t_i}_{12}\Big/
\Big(\bra{\Phi ,t_i}_{12}Q_2^s
\ket{\Phi ,t_i}_{12}\Big)^{1/2}\Big]+$$ $$
\Big(\bra{\Phi ,t_i}_{12}Q_2^{\bar s}
\ket{\Phi ,t_i}_{12}\Big)^{1/2}\Big[
Q_2^{\bar s}\ket{\Phi ,t_i}_{12}\Big/
\Big(\bra{\Phi ,t_i}_{12}Q_2^{\bar s}
\ket{\Phi ,t_i}_{12}\Big)^{1/2}\Big].
\eqno{(6)}$$

When the interaction is completed, the state of the quanton is the reduced density operator $$\rho_1(t_i)=\tr_2
\Big((\ket{\Phi
,t_i}_{12}\bra{\Phi
,t_i}_{12})\Big).
\eqno{(7)}$$

Inserting \$I_2\enskip\Big(=(Q^s_2+Q^{\bar s}_2)\Big)\$ under the partial trace after the bra in (7) (and having in mind idempotency and the above mentioned commutation under the partial trace), it is straightforward  to derive the decomposition of the state \$\rho_1(t_i)\$ of the quanton into its two component states:
$$\rho_1(t_i)=\Big[\tr\Big((\ket{\Phi
,t_i}_{12}\bra{\Phi
,t_i}_{12})Q_2^s\Big)\Big]\rho_1(t_i)^s+
\Big[\tr\Big((\ket{\Phi
,t_i}_{12}\bra{\Phi
,t_i}_{12})Q_2^{\bar s}\Big)\Big]\rho_1(t_i)^{\bar s},\eqno{(8a)}$$
where $$\rho_1(t_i)^s\equiv\tr_2
\Big((\ket{\Phi
,t_i}_{12}\bra{\Phi
,t_i}_{12})Q_2^s\Big)\Big/
\Big[\tr\Big((\ket{\Phi
,t_i}_{12}\bra{\Phi
,t_i}_{12})Q_2^s\Big)\Big],\eqno{(8b)}$$ and
$$\rho_1(t_i)^{\bar s}\equiv\tr_2
\Big((\ket{\Phi
,t_i}_{12}\bra{\Phi
,t_i}_{12})Q_2^{\bar s}\Big)\Big/
\Big[\tr\Big((\ket{\Phi
,t_i}_{12}\bra{\Phi
,t_i}_{12})Q_2^{\bar s}\Big)\Big]\Big\}\eqno{(8c)}$$ are the component states. Naturally, the first-subsystem state decomposition (8a) is actually implied by decomposition (6).

We first \textbf{discuss the preparation in CQM}. Occurrence of the triggering event \$Q_2^s\$ on the screen makes \$\ket{\Phi}_{12}\$ \textbf{collapse} into \$\ket{\Phi}_{12}^c\equiv\ket{\Phi}_{12}^s= Q_2^s\ket{\Phi}_{12}\Big/
\Big(\bra{\Phi}_{12}Q_2^s\ket{\Phi}_{12}
\Big)^{1/2}\$
implying the quanton state \$\rho_1^c=\rho_1^s\$ (cf (8b)). This further implies disappearance of the component \$\ket{\Phi}_{12}^{\bar s}\equiv Q_2^{\bar s}\ket{\Phi}_{12}\Big/
\Big(\bra{\Phi}_{12}Q_2^{\bar s}\ket{\Phi}_{12}\Big)^{1/2}\$
of the state vector of the composite system (the quanton has not hit the rest of the screen) together with the corresponding component \$\rho_1(t_i)^{\bar s}\$ of the state of the quanton (cf (8c)). The surviving component
\$\rho_1(t_i)^s\$ propagates to the measuring instrument.

In \textbf{RSQM} both components \$\ket{\Phi}_{12}^{\bar s}\$ and
\$\ket{\Phi}_{12}^s\$ of \$\ket{\Phi}_{12}\$ evolve in time, each following its own dynamical law. In the former we have some kind of excitation of the 'rest of the screen' and some change of the state of the quanton. But we do not have to follow this up because it is irrelevant for the preparation. In the relative-state approach to \QM it is important to know that this component is believed to exist and evolve in reality though it is not observed in the experiment and not considered in theory.

In RSQM one takes the \textbf{relative state} \$\rho_1^s\$ (cf (8b)) with respect to the event \$Q_2^s\$ on the second subsystem in the composite system state \$\ket{\Phi}_{12}\$ (cf (8b), (4) and (3b)).

Preparation supplies the initial state of the experiment,
hence one is interested only in the further evolution (propagation) of the quanton in the state
\$\rho_1^s\$. It is thus seen that CQM and RSQM  provide us with the same description.\\

{\bf \noindent  4.2a Undetected Triggering
Event with Classical Intuition}

Let us take a purely \textbf{geometrical modification} of one-slit preparation, in which no real occurrence of event, i. e., no detected event takes place in the preparation at \$t_i\$. The geometry is such that if anything is measured on the quanton to the right of the screen at \$t_f\$, \$t_f>t_i\$, classical intuition tells us that the former \textbf{must have passed the slit}, i.e., it is as if the triggering event \$Q_2^s\$ had occurred at \$t_i\$ (cf the preceding version), at the initial moment of the experiment. (This intuitive argument, though it essentially utilizes classical physics, can be reproduced strictly within the \QMl formalism with virtual occurrence \cite{FHRAIO}. It was called retroactive apparent ideal occurrence.)\\

Thus, in \textbf{CQM} one may be tempted to imagine the same physical mechanism as in the preceding version. Namely, classical intuition tells us that passing the slit, i. e. \$Q_2^s\$, must have occurred, and the corresponding collapse must have taken place. This would have had the effect of disappearance of \$\ket{\Phi}_{12}^{\bar s}\$, so that from \$t_i\$ till \$t_f\$ only \$\rho_1^s\$ could exist and evolve  (as far as the quanton is concerned). But in this time interval the \QMl description was in terms of the entire quanton state \$\rho_1\$, given by (8a), because it was not known if the quanton has passed the slit or not.

Here subjectivity in the form of ignorance and false description would seem to creep in. This seems to contradict the fact that \$\rho_1^s=\rho_1^c\$ is part of a pure composite-system state \$\ket{\Phi}_{12}^s\$ (cf (8b) and (3a)), where now \$\ket{\Psi}_{12}^c= \ket{\Psi}_{12}^s\$), and that the pure state \$\ket{\Phi}_{12}^s\$ represents maximal knowledge. This time, unlike in most cases, it does not seem to describe reality. In this attempt of a CQM insight the component \$\rho_1^{\bar s}\$ (cf (8c)), which at \$t_f\$ turns out to not have existed, is the reason for false description in the interval from \$t_i\$ till \$t_f\$.

To my understanding, Bohr would have saved the \CI from these charges leveled against it by forbidding us to consider the interval \$(t_i,t_f)\$ separately. In his opinion it is the entirety of the experiment that matters. (I think, Bohr might have similarly handled the famous Wheeler delayed-choice
experiment \cite{Wheeler}.)\\

In \textbf{RSQM} the description is quite the same as in the preceding version. The entire composite-system state \$\ket{\Phi}_{12}\$ undergoes time evolution from \$t_i\$ till \$t_f\$, and it represents reality. One subjectively chooses \$\rho_1^s\$ at \$t_i\$ as the state relative to the subject event \$Q_2^s\$
in the composite-system state \$\ket{\Phi}_{12}\$. No matter that \$Q_2^s\$ is not "seen" to occur (is not detected); nothing occurs because "occurrence" implies collapse, a notion lacking in this approach.

There is no reason to believe that any detail is false with respect to reality in this approach: the entire state \$\ket{\Phi}_{12}\$ evolves from \$t_i\$ till \$t_f\$, but for the experiment only the mentioned relative state is relevant.

This appears to be a case of clear advantage of RSQM over the expounded CQM version that is burdened with classical intuition.\\

{\bf \noindent  4.2b Undetected Triggering
Event with Discarded Classical Intuition}

The above argument in CQM is not the only possible one. Not knowing at \$t_i\$ if the quanton has passed the slit or not, because the triggering event \$Q_2^s\$ was not detected, one may argue that one should take the entire composite-system state  \$\ket{\Phi}_{12}\$ evolving from \$t_i\$ till \$t_f\$. Only at the latter moment the quanton hits a classical instrument, and collapse takes place \cite{Bohr2}. Then  CQM and RSQM give the same description in the entire interval from \$t_i\$ till \$t_f$.\\

There is a pitfall also in this argument of CQM because one may wonder what is the meaning of the evolution of the part of \$\ket{\Phi }_{12}\$ that determines \$\rho_1^{\bar s}\$ (cf (8c)) when the quanton must have passed the slit. One may be tempted to argue that the evolution of \$\rho_1^{\bar s}\$ must be extinguished retroactively.

Action backwards in time is not in the spirit of \qm . It is preferable to accept the reality of the evolution of \$\rho_1^{\bar s}\$ from \$t_i\$ till \$t_f\$ in CQM just like in RSQM. Then the descriptions coincide, and one has collapse only when occurrence is detected \cite{Bohr2}.\\

One should keep in mind that it is an 'improper mixture' \cite{D'Espagnat}
that is determined by (7), and that the coherence in \$\ket{\Phi}_{12}\$ is not lost. In particular, decomposition (8a), is only apparently an incoherent one, actually it stems from (6), which is a coherent decomposition. Hence physically there is in some sense coherence also between the quanton states \$\rho_1^s\$ and \$\rho_1^{\bar s}\$ (inherited from the decomposition (8a)).

In the one-slit preparation it is hard to think of a way how to observe (bring to interference) the coherence between the two quanton states. Therefore, we discuss next an essentially physically isomorphic experiment that has the advantage of allowing to observe the coherence.\\

{\bf \noindent 5. Mach-Zehnder Interferometer and Wheeler's Delayed-Choice}

To understand the {\bf two
complementary experiments} to be
described, one should have in mind that
the first beam splitter of the Mach-Zehnder interferometer \cite{Bub} can be in
place, can be removed, and can be
replaced by a totally-reflecting mirror
in the same position. When the first beam splitter  is in
place, besides being at the standard
angle \$45^0,\$ it can be at any angle
\$0\leq\theta\leq 180^0\$. Thus, it
plays the role of a preparator. The
photon leaves the preparator at the initial instant \$t_i$.

If the second beam splitter is removed,
we have the \textbf{Mach-Zehnder which-way
device}, and in it the
which-way experiment. If the second
beam splitter is in place, we have the
\textbf{Mach-Zehnder interference device},
and in it the \textbf{complementary experiment},
which is called the interference one.

At the final instant
\$t_f,\$ the photon leaves the place of
the second beam splitter in the former experiment
or the beam splitter itself in the latter one to
enter one of the detectors.\\

If the first beam splitter
is in place at 45 degrees,
we have a coherent initial
state of the photon ($ph$)
$$\ket{ph,t_i}\equiv 2^{-1/2}\Big(
\ket{ph,t_i}^u+\ket{ph,t_i}^l\Big).
\eqno{(9)}$$ Here \$\ket{ph,t_i}^u\$
is a pure photon state starting out to propagate upwards, and after reflection horizontally in the {\it upper arm} towards \$D_H\$, the detector aimed at the upper arm. The state \$\ket{ph,t_i}^l\$ is reflected by the first beam splitter and ready to propagate in the {\it lower arm} horizontally, and then vertically towards \$D_V\$, the detector aimed at the lower arm.\\

{\noindent\bf 5.1 Which-Way with Detected
Triggering Event}

In the \textbf{which-way experiment} the composite (photon plus detectors) system dynamically evolves into the state vector $$\ket{\Phi ,t}_{12}\equiv 2^{-1/2}\Big(
\ket{ph,t}^u_1\ket{D_H}_2+\ket{ph,t}_l
\ket{D_V}_2\Big),\eqno{(10)}$$ where \$t\$, \$t>t_f\$ is the moment of arrival of the photon at the detector (but it is not absorbed yet).

One should note that dynamical evolution has not wiped out the coherence in (9); it has only delocalized it from the photon subsystem to the composite photon-plus-detectors system.

Viewing the which-way experiment in \textbf{\textbf{ CQM}}, one of the detectors is triggered. If it is, e. g., \$D_H\$, it means that the event \$\ket{D_H}_2\bra{D_H}_2\$ occurs, \textbf{collapsing} \$\ket{\Phi ,t}_{12}\$ into \$\ket{ph,t}^u_1\ket{D_H}_2\$, where \$t\$, \$t>t_f\$ is the moment of arrival of the photon at the detector (but it is not absorbed yet). This implies that the photon is in the state \$\ket{ph,t}^u_1\$. If \$D_V\$ is triggered, one reaches the symmetrical conclusion.

In the former case it is clear from the geometry of the instrument that the photon \textbf{had to be transmitted through the first beam splitter }(just like the quanton had to pass the slit in the preceding illustration in the analogous version), and symmetrically in the latter case. Thus, in CQM one actually deals with the mixture $$(1/2)\Big(\ket{ph,t_i}^u\bra{ph,t_i}^u+
\ket{ph,t_i}^l\bra{ph,t_i}^l\Big)\eqno{(11)}$$ (because one does not know which detector will be triggered, but one of them certainly will).

Depending whether \$D_H\$ or \$D_V\$ is eventually hit, the irrelevant  state, \$\ket{ph,t_i}^l\$ or \$\ket{ph,t_i}^u\$ respectively, is effaced. But on account of lack of knowledge which detector will be hit, one has a genuine (not a coherent) mixture in (11).

On the other hand, if we view  the which-way experiment in \textbf{RSQM}, the composite-system state vector (10) does not change at all. We can take the relative state of the photon in \$\ket{\Phi,t}_{12}\$ relative to the state \$\ket{D_H}_2\$ (to the event \$\ket{D_H}_2\bra{D_H}_2\$) of the detector system. Again, one obtains \$\ket{ph,t}^u_1\$ as the photon state. (Analogously in the symmetrical case.) Here we have an illustration for equal \QMl description in CQM and RSQM.

In any of the two approaches we have particle-like behavior of the photon: it seems to propagate only in the upper arm (or only in the lower arm).\\

Wheeler \cite{Wheeler} introduced the following \textbf{delayed-choice} idea. Imagine that the second beam splitter is removed or left in place (which-way or interference experiment respectively) {\bf after} the photon has been prepared on the first beam splitter and before it has reached the place of or the second beam splitter itself. How will the photon behave?

Let \$\bar t\$ (\$t_i<\bar t<t_f\$) be the instant of the delayed decision. Let us consider a sequence of which-way experiments, each equally well described in CQM and in RSQM. Then, suddenly, one switches over in the Wheeler delayed-choice way into a sequence of interference experiments. All photons will end up in \$D_H\$ (interference). Then the CQM description turns out to be wrong because it has an inbuilt collapse at \$t_i\$, which actually has not happened.

In RSQM the initial state (9) of the photon evolves into
$$\ket{ph,t_f}\equiv 2^{-1/2}\Big(
\ket{ph,t_f}^u+\ket{ph,t_f}^l\Big),
\eqno{(12)}$$ and, in the interference experiment this becomes $$\ket{ph,t}_1\ket{D_H}_2,$$ where \$t>t_f\$ as above. (In the which-way experiment the same photon state (12) gives rise to the composite-system state (10).

The description in CQM fares no better when the delayed-choice takes place in the opposite sequences of experiments. Let the first sequence consist of interference experiments. Then the delayed choice suddenly switches over into a sequence of which-way experiments.

In CQM, just like in RSQM, the final state of the photon in the first sequence is given by (10). But the delayed choice makes the second sequence to give $$(1/2)\Big(\ket{ph,t_f}^u_1\bra{ph,t_f}^u_1
\otimes\ket{D_H}_2\bra{D_H}_2\enskip +\enskip
\ket{ph,t_f}^l_1\bra{ph,t_f}^l_1\otimes
\ket{D_V}_2\bra{D_V}_2
\Big),\eqno{(13)}$$ which includes collapse in one of the detectors, and
which necessitates collapse at \$t_i\$ which was not in the description up to the moment \$t\$.

In RSQM in the second, which-way sequence of experiments
the state (10) goes over into $$\ket{ph,t_f}\equiv 2^{-1/2}\Big(
\ket{ph,t_f}^u_1\ket{D_H}_2+\ket{ph,t_f}^l \ket{D_V}_2\Big)$$ (the coherence is always preserved; it is only delocalized from the subsystem to the larger system).\\

It is important and satisfying to know
that both {\it single-photon}
Mach-Zehnder experiments discussed are
no longer in the realm of thought
experiments; they have become real
experiments performed in a convincing
way in the laboratory
\cite{realMZ}.\\

{\bf \noindent 5 Generalizations}\\
The careful reader has noticed that, though we have treated \m , the argument that showed that the ideal collapse and the no-collapse approaches give the same description for subsystem \$1\$ made use only of \$\ket{\Psi}_{12}^f\$ and the event \$Q_2^n\$. Therefore this conclusion can be stated in a broader fashion.

Let us assume that any composite-system state vector \$\ket{\Psi}_{12}\$ and any second-subsystem event (projector) \$Q_2\$ are given. Collapse in terms of ideal occurrence of \$Q_2\$ in \$\ket{\Psi}_{12}\$ on the one hand, and no collapse using \$Q_2\$ as the subject event in the relation $$\rho_1=\tr_2\Big( (\ket{\Psi}_{12}\bra{\Psi}_{12})Q_2\Big)
 \Big/\Big[ \tr\Big( (\ket{\Psi}_{12}\bra{\Psi}_{12})Q_2\Big)\Big].
 \eqno{(14)}$$  for evaluation of the relative state of subsystem \$1\$ give one and the same first-subsystem state and thus \textbf{the same \QMl description} of the object \textbf{in the two approaches }(cf (3b) and (4)).\\

The careful reader has further noticed that (14) does not make sense if \$Q_2\ket{\Psi}_{12}=0\$, or equivalently if \$\bra{\Psi}_{12}Q_2\ket{\Psi}_{12}=0\$, i. e., if the event \$Q_2\$ has zero probability in the state \$\ket{\Psi}_{12}\$. This case must be excluded because \$Q_2\$ cannot occur. One may exclude also the certain event as trivial (no change in the state of the object).\\

\noindent {\bf 6.1} \emph{\textbf{General Occurrence}}

 One suspicion is, no doubt, lingering on in the mind of the reader. Isn't this general coincidence of no-collapse and collapse descriptions a deceptive consequence of the overly idealized and oversimplified notion of ideal occurrence of the subject event \$Q_2\$? We now have to go to occurrence of \$Q_2\$ in a general way.

 For the reader's convenience  an old result of the present author \cite{FH69} (subsection 6(B) there) is reproved in Appendix D. It says that in whatever way \$Q_2\$ occurs , it gives rise to one and the same state of the object subsystem. In other words, all kinds of occurrences give the same as ideal occurrence.

 This conclusion is drawn from the assumption that the probability of the coincidence of two opposite-subsystem events \$P_1\$ and \$P_2\$ in any composite-system state equals the probability that \$P_2\$ occurs, succeeded by an immediate occurrence of \$P_1\$.\\

{\bf \noindent 6.2} \emph{\textbf{General Composite-System State}}
If we have a general (mixed or pure) state \$\rho_{12}\$ instead of the pure state \$\ket{\Psi}_{12}\bra{\Psi}_{12}\$, then the basic relative-state relation (3b), which is general for a pure composite-system state, has to be replaced by a more general relation.

Assuming that we have a proper mixture and arguing in CQM we take resort to the general form of the Von Neumann-L\"{u}ders 'projection' for change of state in ideal \m . It reads
$$\rho_{12}\enskip\rightarrow\enskip
\Big(Q_2\rho_{12}Q_2\Big)\Big/ \Big(\tr(Q_2\rho_{12}Q_2)\Big).\eqno{(15)}$$
It implies $$\rho_1^c=\tr_2\Big(Q_2\rho_{12}Q_2\Big)\Big/ \Big(\tr(Q_2\rho_{12}Q_2)\Big)$$ (generalization of (2)). This can be equivalently rewritten as
$$\rho_1^c=\tr_2\Big(\rho_{12}Q_2\Big)\Big/ \Big(\tr(\rho_{12}Q_2)\Big).\eqno{(16)}$$
(See  how equivalence of (3a) and (3b) is proved.) Finally, we just have to require
\$\rho_1^{rs}\equiv\rho_1^c\$, and we obtain the rhs of (11) as the definition of \$\rho_1^{rs}\$.

Evidently, the subject event \$Q_2\$ has to be restricted by the requirement \$\tr(\rho_{12}Q_2)>0\$, i. e., it has to have positive probability in \$\rho_{12}\$.\\ 
The generalization of this subsection is equally valid for ordinary (first-kind or proper) mixtures \$\rho_{12}\$ as for {\bf improper} (second-kind) ones \cite{D'Espagnat}. In the latter case the two-subsystem state stems from a larger pure state \$\rho_{12}=\tr_{34\dots N}(\ket{\Psi_{12\dots N}}
\bra{\Psi_{12\dots N}})\$ \$N\geq 3\$ (as the reduced density operator). Then the state \$\rho_{12}\$ 'inherits' from \$\ket{\Psi}_{12\dots N}\$ all coherences, i. e., any relevant decomposition of the former into distinct states is due to breaking up the latter into components, which are necessarily coherent (potentially able to cause interference). This is, in some sense,  valid also for \$\rho_{12}\$ as an (apparent) mixture.

If \$\rho_{12}\$ is an improper mixture, we can make use of the fact that one cannot distinguish it physically from a proper mixture described by the same density operator as far as the system under consideration is concerned (system (1+2) in our case). Hence, relation (16) should apply also to improper mixtures.\\

{\bf \noindent 6.3} \emph{\textbf{More than Two Subsystems}}\\
We assume that we have  a general (mixed or pure) state \$\rho_{12\dots N}\$ of a composite system consisting of \$N\$ subsystems (an \$N$-partite system). We want to evaluate the relative state \$\rho_1\$ of the first subsystem, the \textbf{object subsystem}. (It could be any other of the \$N\$ subsystems.) In RSQM we must specify the subject entity. To this purpose we have to select first a subsystem, e. g. subsystem \$2\$ (it could be any other subsystem from the set \{$3,4,\dots ,N$\}).  We call it the \textbf{subject subsystem}. The second and final step is to select an event (projector) \$Q_2\$ for the subject subsystem: the \textbf{subject event}. The subject subsystem and its subject event together constitute the \textbf{subject entity}.

The evaluation of the relative state then goes as follows: $$\rho_1=
\tr_{23\dots N}\Big(\rho_{1\dots N}Q_2\Big)\Big/ \tr\Big(\rho_{1\dots N}Q_2\Big),\eqno{(16a)}$$ or, equivalently (as easily seen) $$\rho_1=
\tr_2\Big(\rho_{12}Q_2\Big)\Big/\tr\Big(\rho_{12}Q_2\Big), \eqno{(16b)}$$
where $$\rho_{12}\equiv\tr_{34\dots N}\rho_{12\dots N}\eqno{(16c)}$$ is the subsystem state (reduced density operator) of the \{$1+2$\} subsystem.\\

{\bf \noindent 7. Concluding Remarks}

The \textbf{basic point} of the entire article is that evaluation of the object state via (16a) (or via (16b) with (16c)) can be \textbf{interpreted in two}, at first glance very different,\textbf{ ways}. One is the way of the \textbf{relative-state approach} claiming that \$\rho_1\$ is the relative state with respect to (or relative to) the subject event \$Q_2\$  in the \$N$-subsystem state \$\rho_{1\dots N}\$ (needless to stress the subject subsystem; the index on the projector does this). The other way is that of any Copenhagen-inspired \textbf{collapse approach}, in which one speaks of the occurrence of the event \$Q_2\$ on subsystem \$2\$ in the same \$N$-subsystem state, but the description applies only to subsystem \$1\$ leaving outside the \QMl description the rest of the subsystems. The occurrence  causes global collapse of the state.

RSQM and CQM are, \textbf{in principle, experimentally distinguishable }as follows. In RSQM \$\ket{\Phi}_{12}\$,
which we take now to be a general
state vector of the \$1+2\$ system, is
unchanged (no collapse). This
implies that subsystem \$2\$ is
described by
\$\rho_2\equiv\tr_1\Big(
\ket{\Phi}_{12}
\bra{\Phi}_{12}\Big)\$, and,
\$\ket{\Phi}_{12}\$ being a
general bipartite state, there may
be entanglement between subsystems \$1\$ and
\$2\$.

Let us take \$Q_2\$ to be an elementary event \$Q_2\equiv\ket{\phi}_2\bra{\phi}_2\$.
In CQM one concludes, according to relation (1), that
subsystem \$2\$ is in the pure
state \$\ket{\phi}_2\$
(after collapse has taken place), and that
there are no correlations between
the states \$\rho_1\$
and \$\ket{\phi}_2\$ of the two
subsystems. (A pure state cannot be correlated as easily seen.)

In this case RSQM and CQM
differ, and the difference is
rather pronounced, and this is, in
principle, experimentally
detectable.

Nevertheless, since collapse is usually
thought to occur on a classical
measuring instrument, it is very
hard to disprove it
experimentally. (At the least, one
would have to find an observable
that is incompatible with the
pointer observable, and measure it.)\\

The mentioned equivalence of the relative-state and of any collapse interpretation can be read from RSQM to CQM saying that any collapse of an event \$Q_2\$ can be understood (physically) as the subjective choice of a subject entity. Contrariwise, reading the equivalence from CQM to RSQM, one can picture the subject entity as a collapse.

The \CI of \QM reigned the physical insight for a long period. Nowadays, there are available very detailed (and rather differing) analyses of it \cite{Copenh}, \cite{Howard}. It appears to the present author that the \CI was \textbf{empirically the simplest way} to make physical sense of quantum experiments in view of the fact that they were performed with classical instruments interacting with \Q objects. The classical instruments were, of course, described by classical physics, where every event was either collapsed (occurred) or so was its opposite event.\\

If one understands the Copenhagen interpretation as a purely \textbf{empirical} one, then there is no point in discarding it. The desirable thing is to explain it. If \textbf{RSQM gives a more fundamental insight}, as this author and many others believe, then, among other things, RSQM  has to derive the Copenhagen interpretation. After the advent of decoherence due to Zeh, this became possible. Namely, decoherence opened the door for understanding how the quantum laws imply the classical worlds as we know it \cite{ZehBOOK}. Then, the above mentioned 'reading the equivalence from left to right' discloses the true physical meaning of the Copenhagen interpretation.

Following the musing of the
Austrian animal behaviorist Konrad
Lorenz, as Tegmark and Wheeler
\cite{TegWheel} (p. 74) write, one
can say that important scientific
discoveries go though three
phases: first they are completely
ignored, then they are violently
attacked as heresy, and finally
they are brushed aside as
prejudice. When the rebellion against the absolute domination of the \CI began, there was a tendency to brush it aside as prejudice. But after the revolutionary ideas of decoherence, I believe that this is no more the case.\\

 Tegmark and Wheeler \cite{TegWheel},  write
in their abstract:

\begin{quote}
"We argue that modern experiments
and the discovery of decoherence
have shifted prevailing quantum
interpretations away from wave
function collapse towards unitary
physics,..."
\end{quote}

 By "unitary physics" one means the assumption that the unitary evolution is never violated, i. e., that there is no collapse. The quoted statement is interesting in view of the facts that Wheeler was Everett's academic adviser at Princeton when the latter wrote his thesis and discovered RSQM. Later Wheeler became sceptical towards RSQM. Near the end of his remarkable life (he died in 2008 at age 96, seven years after the above article appeared in print), Wheeler seems to have returned to RSQM (though this may have been under the influence of his coauthor Tegmark as some private communication suggests).\\

{\bf \noindent Appendix A. Partial Scalar Product}\\
We write arbitrary ket or bra vectors with a bar; those without a bar are unit vectors (as it is in the text). Then a partial scalar product \$\overline{\bra{\phi}_2\ket{\Phi}_{12}}\$ can be evaluated in an arbitrary pair of complete orthonormal bases \$\{\ket{k}_1:\forall k\},\enskip
\{\ket{l}_2:\forall l\}\$. The partial scalar product is obtained {\it in terms of ordinary scalar products} as follows. One expands both factors in the partial scalar product to obtain:

$$\overline{\ket{\phi}}_1\equiv
\overline{\bra{\phi}}_2
\overline{\ket{\Phi}}_{12}=\sum_k\Big\{\sum_l
\Big[\Big(\overline{\bra{\phi}}_2
\ket{l}_2\Big)\Big(\bra{k}_1\bra{l}_2
\overline{\ket{\Phi}}_{12}\Big)\Big]\Big\}
\ket{k}_1.\eqno{(A.1)}$$

The most important point is that the lhs is {\it independent of the choice of the bases} in the tensor-factor spaces. To prove this claim, we take another arbitrary pair of complete orthonormal bases \$\{\ket{p}_1:\forall p\},\enskip\{\ket{q}_2:\forall q\}\$ and we expand the basis vectors in (A.1) in the new bases. In this manner we obtain:

$$lhs=\sum_{p'}\sum_k\Big\{\sum_l
\sum_{p,q,q'}\Big[\Big(\overline{
\bra{\phi}}_2
\ket{q}_2\Big)\Big(\bra{q}_2\ket{l}_2
\Big)\Big(\bra{p}_1\bra{q'}_2
\overline{\ket{\Phi}}_{12}\Big)\times$$ $$\Big(\bra{k}_1\ket{p}_1\Big)\Big(\bra{l}_2
\ket{q'}_2\Big)\Big]\Big\}
\Big(\bra{p'}_1\ket{k}_1\Big)\ket{p'}_1=$$ $$
\sum_{p,p',q,q'}\Big(\overline{\bra{\phi}}_2
\ket{q}_2\Big)\Big\{\sum_l\Big(
\bra{q}_2\ket{l}_2\Big)\Big(\bra{l}_2
\ket{q'}_2\Big)\Big\}\times$$ $$
\Big\{\sum_k\Big(\bra{p'}_1\ket{k}_1\Big)
\Big(\bra{k}_1\ket{p}_1\Big)\Big\}
\Big(\bra{p}_1\bra{q'}_2
\overline{\ket{\Phi}}_{12}\Big)\ket{p'}_1.$$

The large brackets in the last expression, reading from left to right, are \$\delta_{q,q'}\$ and \$\delta_{p',p}\$ respectively. Taking this into account, one further obtains:

$$lhs=\sum_p\Big\{\sum_q
\Big[\Big(\overline{\bra{\phi}}_2\ket{q}_2
\Big)
\Big(\bra{p}_1\bra{q}_2\overline{
\ket{\Phi}}_{12}\Big)\Big]\Big\}\ket{p}_1.
\eqno{(A.2)}$$

Comparing (A.1) and (A.2), one can see that they are of the same form. If one started out with the second pair of bases, one would obtain the rhs of (A.2). And this equals \$lhs\equiv\overline{\ket{\phi}_1}\equiv
\overline{\bra{\phi}_2}
\overline{\ket{\Phi}_{12}}\$ as seen from (A.2). This concludes the proof.\\

{\it Remark A.1} Naturally,  the partial scalar product \$\overline{\bra{\phi}_2}
\overline{\ket{\Phi}_{12}}\$ ca be evaluated also by expressing \$\overline{\ket{\Phi}_{12}}\$ in any other way as a finite or infinite linear combination  of tensor products of tensor-factor vectors.

{\it Remark A.2} The partial scalar  product, like the full (or ordinary) scalar product, has the properties of linearity and continuity. Making use of these properties, one can arrive at the last expression in (A.1)  expanding only \$\overline{\ket{\Phi}}_{12}\$ (without the expansion of \$\overline{\bra{\phi}}_2\$).\\

\vspace{0.5cm}

{\bf\noindent  Appendix B. Proof of Equivalence of the Partial-Scalar-Product\\ and the Partial-Trace Relations}\\

\noindent
Let us start with the partial-scalar-product relation:
$$\ket{\phi}_1\bra{\phi}_1=
\Big(\bra{\phi}_2\ket{\Phi}_{12}\Big)\Big(\bra{\Phi}_{12}\ket{\phi}_2
\Big)\Big/\Big(\bra{\Phi}_{12}(I_1\otimes\ket{\phi}_2
\bra{\phi}_2)\ket{\Phi}_{12}\Big).
\eqno{(B.1)}$$  First we evaluate the operator \$\cN_1\$ that is the nominator on the rhs utilizing two arbitrary complete orthonormal bases in the tensor-factor spaces \$\{\ket{i}_1:\forall i\},\enskip\{\ket{n}_2:\forall n\}\$.

$$\bra{i}_1\cN_1\ket{i'}_1=
\Big(\sum_n\bra{\phi}_2
\ket{n}_2\bra{i}_1
\bra{n}_2\ket{\Phi}_{12}\Big)
\Big(\sum_{n'}\bra{\Phi}_{12}\ket{i'}_1
\ket{n'}_2\bra{n'}_2
\ket{\phi}_2\Big).\eqno{(B.2)}$$

On the other hand, let us take the operator \$\cN_1'\$ that is the nominator on the rhs of the partial-trace relation (cf (3b) with \$\ket{\Psi}_{12}\$ etc. replaced by \$\ket{\Phi}_{12}\$ etc and \$\ket{\phi}_{12}\bra{\phi}_{12}\$ substituted for \$Q_2\$):
 $$\ket{\phi}_1\bra{\phi}_1=\tr_2
 \Big[\Big(\ket{\phi}_2\bra{\phi}_2\Big) \Big (\ket{\Phi}_{12}\bra{\Phi}_{12})\Big)\Big]
 \Big/ \tr\Big[\Big(\ket{\phi}_2
 \bra{\phi}_2\Big) \Big (\ket{\Phi}_{12}\bra{\Phi}_{12})\Big)
 \Big].\eqno{(B.3)}$$ Let us evaluate (B.3) in the same two bases. We obtain
 $$\bra{i}_1\cN_1'\ket{i'}_1=\sum_{n'}
 \sum_n\bra{n'}_2\ket{\phi}_2
 \bra{\phi}_2\ket{n}_2
 \bra{i}_1\bra{n}_2\ket{\Phi}_{12}
 \bra{\Phi}_{12}\ket{i'}_1
 \ket{n'}_2.\eqno{(B.4)}$$

 Obviously, the terms on the rhs of (B.2) and (B.4) are equal (product of the same four numbers in different order). Hence, \$\cN_1=\cN_1'\$.

 Next, we turn to the numbers that are the denominators \$\cD\$ and \$\cD'\$ of the partial-scalar-product and the partial-trace relations respectively,  and we ascertain of their equality.

 $$\cD =\sum_{i,n,i',n'}\Big[\bra{\Phi}_{12}
 \ket{i}_1\ket{n}_2\Big]
 \Big[\bra{i}_1\bra{n}_2
 \Big(I_1\otimes \ket{\phi}_2\bra{\phi}_2\Big)\ket{i'}_1
 \ket{n'}_2\Big]\times$$ $$
 \Big[\bra{i'}_1\bra{n'}_2\ket{\Phi}_{12}
 \Big]
 =\sum_{i,n,i',n'}\bra{\Phi}_{12}
 \ket{i}_1\ket{n}_2\delta_{i,i'}\bra{n}_2
 \ket{\phi}_2\bra{\phi}_2
 \ket{n'}_2\times$$ $$
 \bra{i'}_1\bra{n'}_2
 \ket{\Phi}_{12}.\eqno{(B.5)}$$

 On the other hand we have $$\cD'=\sum_{i,n,i',n'}\delta_{i,i'}
 \bra{n}_2\ket{\phi}_2\bra{\phi}_2
 \ket{n'}_2\Big[\bra{i'}_1\bra{n'}_2
 \ket{\Phi}_{12}\Big]\Big[
 \bra{\Phi}_{12}\ket{i}_1 \ket{n}_2\Big].\eqno{(B.6)}$$

 Again we can see that the corresponding terms in (B.5) and (B.6) coincide. This concludes the proof.

 {\it Remark} It is not necessary to prove that the above partial trace is basis independent because this has been proved for the partial scalar product in Appendix A, and now equivalence of the partial trace and the partial scalar product has been established.\\

{\bf \noindent APPENDIX C. The general pure-composite-state relative-state relation and Everett.}\\
Let a pure state \$\ket{\Psi}_{12}\$ of a composite system be given, and
let us assume that the event \$Q_2\$ is an elementary one (mathematically a ray projector) \$Q_2=\ket{\phi}_2\bra{\phi}_2\$. The basic pure-state relative-state formula (3b) then determines the relative state $$\rho_1=\tr_2\Big(\ket{\Psi}_{12}
\bra{\Psi}_{12}\ket{\phi}_2\bra{\phi}_2
\Big)\Big/\tr\Big(\ket{\Psi}_{12}
\bra{\Psi}_{12}\ket{\phi}_2\bra{\phi}_2
\Big)\Big.
\eqno{(C.1)}$$ Rewriting this equivalently in terms of a partial scalar product (cf Appendix A and relation (B.1)), the relations
$$\rho_1=\ket{\psi}_1\bra{\psi}_1,\quad
\ket{\psi}_1=\bra{\phi}_2\ket{\Psi}_{12}
\Big/\Big[\bra{\Psi}_{12}\Big(I_1\otimes
\ket{\phi}_2\bra{\phi}_2\Big)\ket{\Psi}_{12}
\Big]^{1/2}\eqno{(C.2)}$$ ensue.

Let us take an ortho-normal complete basis in the state space of the second subsystem such that \$\ket{\phi}_2\$ is one, say the first, of the basis vectors: \$\ket{\phi}_2+\{\ket{i}_2:i=2,3,\dots\}\$.
Expanding \$\ket{\Psi}_{12}\$ in this subsystem basis, one obtains $$\ket{\Psi}_{12}=\overline{\ket{\psi}_1}
\ket{\phi}_2+\sum_i\overline{\ket{i}_1}
\ket{i}_2,\eqno{(C.3a)}$$ and $$\overline{\ket{\psi}_1}=\bra{\phi}_2
\ket{\Psi}_{12},\quad\overline{\ket{i}_1} =\bra{i}_2\ket{\Psi}_{12},\enskip i=2,3,\dots\eqno{(C.3b)}$$

Then (C.2) gives $$\ket{\psi}_1= \overline{\ket{\psi}_1}\Big/ ||\overline{\ket{\psi}_1}||.\eqno{)C.4)}$$

Since a ray projector, like \$\ket{\phi}_2\bra{\phi}_2\$, can in \QM be interpreted both as an event and as a pure state, one can decide upon the latter and say with Everett \cite{Everett1}, \cite{Everett2} that \$\ket{\psi}_1\$ {\bf is the relative state of the object subsystem with respect to the state \$\ket{\phi}_2\$ of the subject subsystem}.\\

{\bf \noindent Appendix D. Arbitrary Occurrence of \$Q_2\$}

If  \$\rho_{12}\$ is the state (density operator) of a composite system and a subsystem event \$P_2\$ occurs in it, the probability of occurrence is \$\tr\Big(\rho_{12}P_2\Big)\$. The state changes in an unknown way (depending on the way how \$P_2\$ occurs) into some state \$\rho'_{12}\$. The probability of an immediately subsequent occurrence of an opposite-subsystem event \$P_1\$ in the new state is \$\tr\Big(\rho'_1P_1\Big)\$, where \$\rho'_1\equiv\tr_2\rho'_{12}\$ is the state (reduced density operator) of the first subsystem. The total probability that both \$P_2\$ and \$P_1\$ occur in immediate succession in \$\rho_{12}\$ is
$$\tr\Big(\rho_{12}P_2\Big)\times \tr\Big(\rho'_1P_1\Big).\eqno{(D.1)}$$

On the other hand, if one measures the same two events in coincidence in the same state \$\rho_{12}\$, then the probability is
$$\tr\Big(\rho_{12}P_1P_2\Big)=$$ $$\tr\Big[P_1\tr_2\Big(\rho_{12}P_2\Big)\Big]=
\tr\Big(\rho_{12}P_2\Big)\times\tr\Big\{P_1 \Big[\Big(\tr_2(P_2\rho_{12}P_2)\Big)\Big/
\Big(\tr\rho_{12}P_2\Big)\Big]\Big\}.\eqno{(D.2)}$$ One should note that total traces are written without indices. They are taken in the  state space where the operator under the trace acts. Besides, we have again utilized the idempotency of the projector and commutation under the partial trace.

The 'immediate succession' of occurrence of the events in (D.1) means the following. If \$\Delta t\$ is the time lag between the two occurrences, one can take \$limes\Delta t=0\$. Since the expressions are continuous and the events are compatible (commuting projectors), it is plausible to assume that, in spite of the fact that \$\rho'_{12}\$ is unknown, the probability of the succession of occurrences (D.1) converges into the probability of the coincidence of occurrences (D.2). Then, equating (D.1) with the rhs of (D.2),  \$\rho_1'\$ becomes equal to the L\"{u}ders (selectively) changed state
\$\Big(\tr_2(P_2\rho_{12}P_2)\Big)\Big/
\Big(\tr(\rho_{12}P_2)\Big)\$ as claimed.\\


ACKNOWLEDGEMENT This investigation is an elaboration of  an inspiration due to a remark from Dieter Zeh in private e-mail communication. I am thankful for the latter.

\end{document}